# The Neutral Hydrogen Disk of Arp 10 (= VV 362) : A Non-equilibrium Disk Associated with a Galaxy with Rings and Ripples

V. Charmandaris and P. N. Appleton

*Department of Physics and Astronomy, Iowa State University, Ames, IA 50011*

## ABSTRACT

We present VLA[1] H I and optical spectra of the peculiar galaxy Arp 10. Originally believed to be an example of a classical colliding ring galaxy with multiple rings, the new observations show a large disturbed neutral hydrogen disk extending 2.7 times the radius of the bright optical ring. We also present evidence for optical shells or ripples in the outer isophotes of the galaxy reminiscent of the ripples seen in some early type systems. The small elliptical originally believed to be the companion is shown to be a background galaxy. The H I disk consists of two main parts: a very irregular outer structure, and a more regular inner disk associated with the main bright optical ring. In both cases, the H I structures do not exactly trace the optical morphology. In the outer parts, the H I distribution does not correlate well with the optical ripples. Even the inner H I disk does not correspond well morphologically nor kinematically to the optical rings. These peculiarities lead us to believe that the potential in which the H I disk resides is significantly out of equilibrium — a situation which would inherently produce rings of star formation. We suggest that Arp 10 is the result of the intermediate stage of a merger between a large H I rich disk and a gas-poor disk system. As such, it may represent an example of a class of mergers which lies intermediate between the "ripple and shell" accretion systems and the head-on collisional ring galaxies.

---

astro-ph/9510003   02 Oct 1995

## 1. Introduction

Arp 10 (= VV 362) is a galaxy containing a bright ring, an off-center nucleus and a faint bar. These features can be easily detected in the photograph presented in the Arp Atlas of Peculiar Galaxies (Arp 1966), suggesting that Arp 10 may be a colliding ring system. This galaxy was chosen as part of a larger study of the dynamics and star formation properties of ring systems (See Appleton and Struck-Marcell, 1995 for review).

Recent deep CCD imaging of the galaxy in H$\alpha$ and R–band by Charmandaris, Appleton and Marston 1993, hereafter CAM, revealed, in addition to the bright ring, an very small inner ring and traces of outer ring arcs. The earlier observations also provided evidence for threshold behavior of the star formation rate along the rings. In that paper we suggested that Arp 10 may be an example of a collisional ring galaxy, in which rings are produced as a result of the passage of a small galaxy through the center of a larger rotating disk (Lynds and Toomre 1976).

Deep broad-band imaging by Appleton and Marston (1995) now show that, in addition to the bright inner rings, the galaxy exhibits faint optical structure reminiscent of shells or "ripples" seen around some early type galaxies (e.g. Schweizer and Seitzer 1988; hereafter SS88). A reproduction of the B–band image of Arp 10 is shown in Figure 1. The fact that Arp 10 shows both "ripples" and rings suggests a formation history which may be intermediate between a classical ring galaxy and a merger/mass transfer event which is thought to responsible for most "ripples" or shells (Quinn 1984, Dupraz & Combes 1986, Hernquist and Quinn 1987). The discovery of an extra-nuclear knot by CAM 5$''$ to the southwest of the nucleus of Arp 10 (see also Figure 9 of this paper) lends further support to the idea that Arp 10 is some form of merging system, since the knot might be the nucleus of a second galaxy. The study of the morphology and kinematics of the H I in this system is therefore of considerable interest in the search for a complete explanation for the formation of rings and shells, but also for the ultimate fate of gas in such systems.

The single dish H I spectrum of the galaxy obtained by Sulentic and Arp (1983; hereafter SA) using the Arecibo radio telescope, exhibits the typical two-horned profile of a rotating planar disk. It was assumed that such a disk would be associated with the bright optical ring.

In order to further study the kinematics of the system we obtained medium resolution H I observations of Arp 10 using the C-array of the Very Large Array. We present a detailed mapping of the galaxy and provide a more complete picture of the internal kinematics of its gaseous content. In § 2 we describe our observations and in § 3 we present the global H I characteristics of the system. In § 4 we elaborate on the H I distribution and the kinematics of the galaxy. In § 5 we present the longslit spectral observations and in § 6 we examine the plausible scenarios that could lead to the formation of this system. Finally, in § 7 we present our conclusions.

We assume throughout this paper a distance to Arp 10 of 121 Mpc, based on a heliocentric velocity for Arp 10 of 9108 km s$^{-1}$ (this paper) and a Hubble constant of 75 km s$^{-1}$ Mpc$^{-1}$.

## 2. Observations

The observations were made on December 4 1994, using all 27 telescopes in the C-array of the VLA. We used a bandwidth of 3.125 MHz centered at 9,104 km s$^{-1}$, the heliocentric optical velocity of the galaxy. The correlator was set in the 2AC mode with on-line hanning smoothing and 64 channels. This provided a frequency separation of 48.8 kHz per channel, which corresponds to 10.94 km s$^{-1}$ in the rest-frame of the galaxy using the optical definition of redshift. The velocity coverage of our observations was 689 km s$^{-1}$. A total of 4 hours 53 min was spent on source. Flux and phase calibration was achieved using the sources 3C48 and 0202+149 (1950) respectively.

These data were first amplitude and phase calibrated and bad data due to interference were flagged and ignored by the AIPS software. Two separate image cubes were created from the UV data using the AIPS task HORUS. The first image cube (hereafter cube-1), was created by giving more weight to those baselines which sampled the uv plane more frequently (so-called natural weighting). This provided a synthesized beam with a FWHM of 21.5$''$ × 20.2$''$. For the second (hereafter cube-2) we used the uniform weighting scheme, which gives equal weight to every sampled uv datum. This provided a smaller beam (FWHM of 13.8$''$ × 12.2$''$) but was less sensitive to extended emission. In this paper we will present the results from cube-1. We analyzed, for completeness, the data for cube-2 and the results were consistent with those derived from the study of cube-1. An inspection of the channel maps of cube-1, showed that there was H I emission from 27 channels covering the velocity range 8,961 to 9,246 km s$^{-1}$.

The subtraction of the continuum emission in each line map was performed using a standard interpolation procedure based on nine continuum maps free from H I at each end of the band. The channel maps were corrected for the effects of the sidelobes of the VLA using the CLEAN procedure described by Hogbom (1974), down to a level of 1.5 times the rms noise. The resulting delta functions were restored with a Gaussian shaped beam of dimensions 21.0″ × 21.0″ for cube-1 and 13.0″ × 13.0″ for cube-2.

In order to determine the total H I distribution we used the following technique on cube-1 which was the most sensitive in extended emission. Initially, we smoothed all channel maps to a resolution 42.0″ × 42.0″. New maps were formed comparing, pixel for pixel, the original full resolution maps to the smoothed ones. The pixel values of original maps were copied to the new ones only if the signal-to-noise ratio of the smoothed map at that point exceeded 2. The total H I surface density map was produced by adding the new maps all together. The same technique was used to create the first and second moment maps of the distribution. Using this procedure we effectively give more weight to points associated with low surface brightness emission.

Optical spectra of Arp 10 were obtained on the nights of 28 and 29 of November 1992. Further spectra were made of the the apparent elliptical companion companion by P. N. Appleton and C. Winrich (Winrich and Appleton 1995) under non-photometric conditions on the night of January 30 1995. All optical spectra were obtained using the Goldcam spectrograph at the KPNO 2.1 m telescope and a full description of these observations will be presented elsewhere. The spatial resolution of the spectra was 1.56 arcsec pixel$^{-1}$ and the dispersion 1.52 Å pixel$^{-1}$.

## 3. Global Profile and H I Distribution

The integrated H I profile of Arp 10, is presented in Figure 2. The two-horned profile of a rotating disk is well defined. The integrated flux density $\int S(V)dV$ detected by SA using Arecibo was 3.1 Jy km s$^{-1}$. Our observations detected 2.75 Jy km s$^{-1}$ which accounts for 89% of the total emission quoted by SA. There is a small asymmetry in our profile with the left (low velocity) horn broader and with smaller peak flux value than the right (high velocity) one. This is opposite to the one appearing in the spectrum of the single dish observations (Figure 1 of SA). These differences, though small, may relate to the fact that the C-array is not sensitive to all the flux in the source. We are clearly missing approximately 10% of the flux seen in the single dish data. On the other hand, small differences in the pointing of the Arecibo telescope relative to the kinematic center of the galaxy could account for the differences in the line shapes.

Assuming that the gas is optically thin, the total H I mass $M_H$, can be calculated using the relation,

$$M_H/M_\odot = F_H D^2 \qquad (1)$$

where $F_H$ is

$$F_H = 2.356 \times 10^5 \int S(V)dV = 6.48 \times 10^5 \, Jy\,km\,s^{-1} \qquad (2)$$

and, $D$ is the distance in Mpc, $S(V)$ is the flux density in Jy and $V$ is the velocity in km s$^{-1}$. The total H I mass of Arp 10 detected with the C-array is $M_H = 9.5\,10^9 M_\odot$

We present in Table 1 the derived H I properties of Arp 10. The heliocentric velocity of Arp 10, 9108 km s$^{-1}$, is in close agreement with the value obtained by SA of 9093 km s$^{-1}$.

The integrated H I distribution of Arp 10 is presented in Figure 3. We observe that the hydrogen emission is very extended and it is distributed over a roughly elliptically-shaped area (see Table 1). The H I dimensions are 2.7 times the diameter of the bright ring seen in the Arp (1966) Atlas (Ring 2 of CAM) and extends well outside the dimensions of the faint optical "ripples" seen in Figure 1. The H I emission exhibits strong peaks in two areas, one southeast and northwest of the nucleus. The northwestern peak seems to coincide with the part of Ring 2 which also displays strong Hα emission (CAM).

## 4. The kinematics of Arp 10

We will show in this section that the kinematic behavior of the gas in Arp 10 is rather complex. We can begin to appreciate the complexity by inspection of the sequence of channel maps shown in Figure 4. Each map represents the H I surface density observed over a single channel of velocity width of 10.8 km s$^{-1}$.

Unlike the channel maps presented for normal galaxies (see for example the work of Wevers, 1984),

the results in Figure 4 show a complex behavior indicating a disturbed or possibly warped system. We summarized the content of the channel maps in Figures 5a and b by splitting the emission features seen in the channel maps into two parts. Those which occur principally in the OUTER regions of the galaxy are shown in Fig. 5a and those which are associated with the INNER disk and ring in Fig. 5b. The dark lines indicate the approximate extent of the H I emission ridge-lines or centroids in each channel map superimposed on a deep B-band image of the galaxy (See Appleton and Marston 1995 for further details of the optical observations).

### 4.1. The Outer H I Structure

We will begin by discussing the outer H I structures in Figure 5a (Color Plate). An interesting aspect of the outer H I structures is the degree to which the H I structure fails to correlate with the faint optical "ripples" seen in the outer regions. The highest velocity emission, around 9350–9150 $\mathrm{km\,s^{-1}}$, develops into a major horseshoe-like loop [2] (seen especially at velocities around 9170–9192 $\mathrm{km\,s^{-1}}$). This H I structure feature extends much further south than any of the faint optical loops or filaments. As we proceed to lower velocities (9115–9000 $\mathrm{km\,s^{-1}}$) this outer structure breaks up into knots and arclets of emission to the west of Arp 10. The overall impression is that of an irregular outer H I disk which does not show any kinematic peculiarities which correlate strongly with wisps of faint optical emission.

There is considerably more coherence to the run of position with velocity on the eastern side of the galaxy were we see one of the few H I/optical correlations of the entire system. This occurs in the velocity range 9170 to 9016 $\mathrm{km\,s^{-1}}$. In this interval the H I centroids at each velocity follow very closely the faint "arm-like" structure which protrudes from the southeastern end of the bright ring. The velocity gradient along this feature is quite constant (around 5.6 $\mathrm{km\,s^{-1}\,kpc^{-1}}$) . As one proceeds to lower velocities, emission is seen over a considerable region around the northwestern end of the ring. The most striking aspect of the H I structure is seen at the lowest velocities when loop-like structures are again seen in the channel maps, almost mirroring the high velocity emission described earlier. These loops, seen at velocities of between 8995–8962 $\mathrm{km\,s^{-1}}$, again have no direct optical counterparts. Unlike the high-velocity "horseshoe", these loops cannot be ascribed as an artifact of a falling rotation curve. One end of these kinematic loops connects to eastern end to the optical "arm" giving the appearance of a giant shell in the channel maps. Our main point here is to show that, although the outer H I disk of Arp 10 exhibits no obvious enhancements in H I surface density near optical features, there is evidence in the northern disk for correlations between kinematic features and the peculiar optical "arm". This is important because it suggests that at least some parts of the outer H I disk may be causally connected to the faint outer optical emission.

### 4.2. The Inner Regions of Arp 10 near the Bright Optical Ring

Are there better correlations between H I features and optical structures in the inner regions of Arp 10 ? In order to investigate this, we present in Figure 5b the inner H I ridge-lines and emission centroids superimposed on a grey-scale representation of the bright inner regions of the optical galaxy. Firstly, as was obvious from the integrated H I map presented earlier, there is strong H I associated with the northern bright region of the optical ring. H I emission can be traced from around 9006 $\mathrm{km\,s^{-1}}$ at the northwestern end of the ring around both sides of the ring to about 9071 $\mathrm{km\,s^{-1}}$ which follow approximately the optical ring. However, as we proceed to higher velocities, the spatial coincidence of the H I with the optical ring breaks down. The emission centroids along the western edge of the ring cross over the optical ring and are seen projected on the inside of the ring at velocities of 9170 $\mathrm{km\,s^{-1}}$. On the eastern side there is almost no correlation with ring position. In fact very little H I is seen distinctly associated with the southern end of the ring (this corresponds to a marked depression in the integrated H I map of Figure 3). The "arm-like" structure which was discussed above, is seen to develop to the east of the southern end of the ring but seems quite separate from the ring itself and shows no coherence with the ring. In one channel, an anomalous H I cloud is seen in the southwest quadrant of the ring at a velocity of 9060 km s$^{-1}$.

---

[2] We note that the appearance of the horseshoe loop in the channel maps at these extreme velocities is a symptom of the apparently falling rotation curve of Arp 10 in its southern quadrant. In a simple disk with a falling rotation curve, features similar to horseshoe (but more closed at one end) would be expected. However, in this case, because of differences between the inner and outer disks, the horseshoe shape is much more open than would be normally expected giving it the appearance of a partial ring or loop.

Given the loopy and peculiar nature of the H I features traced in the channel maps, and the lack of obvious correlation between the H I and optical structures (except for the northern end of the bright ring and the eastern "arm"), it is perhaps surprising to find that the overall velocity field of the system looks, superficially, like a normal rotating disk. The mean velocity field of the H I emission is shown in Figure 6. The overall impression is that of a single coherent rotating disk of H I with a kinematic major axis close to a PA of 0°. The northern end of kinematic major axis shows contours of increasing velocity with radius, symptomatic of a solid-body rotation curve, whereas the southern end of the galaxy shows a turn-over on the peak velocity and a slow decline thereafter with radius. A close inspection of Figure 6 will show that deviations from this simple picture are apparent at the approximate radius of the bright ring and in the outer regions of the disk, especially to the NE and the SW. The magnitude of the deviations are approximately 30–50 km s$^{-1}$. It is clear from our earlier discussion of Figures 5a and 5b that the reason for the change in the isovelocity field at this radius is due to the shift in emphasis from the loopy outer structures, which have an approximate north-south kinematic axis, to an inner structure dominated by emission from the northern end of the ring and the inner stem of the eastern "arm". However, as Figures 5a,b show, the emission in both regions is far from normal for a simple rotating disk.

A velocity dispersion map of the H I was constructed but was not found to be helpful in the interpretation of the kinematics of Arp 10 and is not shown here. The map showed that the highest velocity dispersion in the H I was 76 km s$^{-1}$ and it was located 5″ north of the nucleus.

### 4.3. Attempts to fit a simple disk model

We devoted an extensive amount of effort in order try to explain the overall kinematics of the H I in terms of a) a simple rotating disk (with possible tilted rings), or b) a set of rotating and expanding rings of material, as might be expected from a collisional ring galaxy model (see Appleton and Struck-Marcell 1987 for example). In all cases these simple models failed to provide a good description of the H I kinematics. We outline below these unsuccessful attempts and argue that the failure to model the system is indicative of a disk which is extremely disturbed.

- Our first attempt was to fit a set of rotating tilted rings of increasing radius to the galaxy mean velocity field. Here we used the AIPS task GAL. The disturbed morphology of the northern region of the velocity field made it impossible to use a single model for the whole galaxy. Deviations of the order of 50 to 100 km s$^{-1}$ were found if both the northern and southern ends of the galaxy were included. The fit was only moderately acceptable when we restricted the fitting to a wedge area between PA=140° and PA=200°. We note that the position angle for the kinematic axis in this case was found to be 175° which is significantly different from the major axis of the ring systems (130°). We conclude that a simple model is only a very rough approximation to the overall kinematics of the disk.

- The second attempt was to explore a model which assumes that at each radius the disk may be both rotating and radially expanding. Such a model has been successful in explaining the kinematics of the bright star forming ring in the Cartwheel ring galaxy (Fosbury and Hawarden 1977; Higdon 1993). Assuming reasonable values for the eccentricity and inclination of the rings we calculated the velocity $v$ as a function of the deprojected azimuth $\phi$, along a series of 14 elliptical rings of increasing radius. The position angle of the major axis of the rings was forced to be identical to the optical rings (i.e. PA=130°) Then, we performed a three parameter fit to each ring using the following function:

$$v = a_0 + a_1 sin(\phi + \phi_0) \qquad (3)$$

where $a_0, a_1$ and $\phi_0$ were the free parameters. A rotating and expanding ring would exhibit a simple sinusoidal shape, whose phase offset from the major axis is related to the amplitude of the expansion. The fits were generally poor, with calculated $\chi^2$ per degree of freedom for each ring ranging from 0.39 to 0.53. We used a range of values for the eccentricity and the inclination of the rings, but no satisfactory fit was found. We concluded that velocity asymmetries of the order of 30–50 km s$^{-1}$ were the primary cause of the failure of this approach.

The above results reinforce our belief that Arp 10 is a more complex collisional system that it was originally thought.

## 4.4. A Comparison between the Optical and H I Velocities

As discussed in Section 2 we obtained optical spectra along three slit positions with the KPNO 2.1m telescope. Two were through the center of the galaxy, along the major and minor axes of the rings, and a third was parallel to the minor axis but was offset to the extremely bright H II regions in the northwestern quadrant of the ring.

In Figures 7a and b, we show position-velocity diagrams through the center of the galaxy along both the major and minor axes of the rings (major axis assumed to be at PA=130°). The solid lines show the H I velocities (derived by taking slices through the H I velocity field at the appropriate positions of the optical slits) and the black dots show velocities measured in the H II regions using our KPNO spectra (derived from the H$\alpha$ line). Within the errors of the optical observations (30 kms$^{-1}$), there is generally good correspondence between the optical and H I velocities for the major axis slice. There is a slight suggestion that the optical velocities are systematically higher by 50 kms$^{-1}$ than those obtained in the H I. Our average optical velocity for the galaxy is 9160 kms$^{-1}$ compared with the H I velocity of 9108 kms$^{-1}$. This velocity difference could be explained by a slight misalignment between the slit position and the nucleus during the optical observations, or alternatively could reflect a real difference between the velocity of the optical nucleus and the H I disk.

The asymmetry in the shape of the rotation curve is obvious. The northwestern disk shows a rising rotation curve, whereas the southeastern side of the galaxy flattens out at a radius of about 10 arcsecs. The position-velocity slice along the minor axis (Figure 7b) is peculiar in that large velocity gradients are observed (of order 150 kms$^{-1}$ over an angular scale of 10″). The optical velocities in the southwest quadrant of the galaxy seem to be higher than those found in the H I, even taking into account the systematic effect discussed above. Higher resolution spectra taken along different position angles in the system would be highly desirable to confirm these apparent differences.

## 5. The Search for a Nearby Companion to Arp 10

In our earlier purely optical study of the photometric properties of Arp 10 (CAM), we presented evidence of collisionally induced star formation in the galaxy. Based on the optical appearance of the bright regions of Arp 10 we had hypothesized that the dwarf elliptical galaxy located 60″ northeast of the nucleus of Arp 10 was the "intruder" galaxy responsible for the formation of the ring structure by the process discovered by Lynds and Toomre (1976). However, optical longslit spectral observations of the elliptical galaxy presented here show that the elliptical is a background galaxy. As one can see in Figure 8, the spectrum of the elliptical galaxy clearly shows the Ca II H and K lines, and the G–band in absorption. The estimated velocity of the galaxy based on these lines is 26,680 kms$^{-1}$ (z=0.089). This contrasts sharply with Arp 10 which has a redshift of z=0.03, indicating that the elliptical galaxy is not the companion to Arp 10. However, we will argue below that the presence of shells or "ripples" in the outer regions of the galaxy may be evidence for the disruption of a companion which may be in the process of merger.

## 6. Arp 10: Disk Formation or a Severely Disturbed Disk?

Arp 10 is not a normal galaxy. Optically, its appearance alone placed it in the peculiar galaxy catalogs of Arp (1966) and Vorontsov-Velyaminov (1977). If we consider the spatial distribution and kinematics of the H I alone, without regard to the optical image, we would conclude that Arp 10 consists of a large, quite disturbed, rotating gas disk. Because of our inability to fit a simple rotating disk model to the system and the fact that it appears, superficially at least, to have no nearby companion to stir the disk, it is natural to ask whether this could be an example of a primordial H I cloud which is just organizing itself into a disk system. This argument would appear to break down when the optical galaxy is considered. Broadband photometry (Appleton and Marston 1995) of the galaxy show it to be one of the reddest ring systems in a sample of 12 such galaxies studied ($B-V$=0.8), suggesting a substantial old stellar population. The existence of a weak bar, star forming rings and shells in the outer parts of the galaxy suggest a collisional interpretation for the disturbed H I kinematics. We will therefore not consider the primordial hypothesis further.

The discovery of faint loops and ripples in the outer optical isophotes suggests that the system may have much in common with the "ripples around disk" systems of SS88. These authors discovered that, like the shells around elliptical galaxies (Malin and Carter 1980), early-type disk galaxies can also be surrounded by shells or ripples of faint optical emission. Schweizer and Seitzer argued convincingly that the ripples were external in origin and probably were the result of debris from a major accretion event onto a disk from a nearby companion. The implication from their work is that significant mass transfer or even mergers involving a low-mass companion may not always lead to the complete disruption of the larger disk system. As we discussed in Section 4, Arp 10 also has a loopy outer H I distribution. Recent work by Schiminovich et al. (1994, 1995) has shown that weakly correlated H I/optical structures have been seen in shell elliptical galaxies.

We believe that the evidence is strongly in favor of Arp 10 being primarily a disk system which is in a non-equilibrium state. The existence of relatively large non-circular motions in the H I disk, combined with the very unusual distribution of the H I loops and filaments and their mismatch with the equally anomalous optical "ripples" is clear evidence for a disturbed system. **The overall kinematics of the large H I disk suggests a system in which rotation dominates over radial motions, but only just**. Various parts of the galaxy are experiencing quite large non-circular motions of the order of 30 to 50 $km\,s^{-1}$ (approximately 25–30% of the rotational velocities). The discovery by CAM of three successively nested rings or ring-arcs of star formation in Arp 10 was interpreted by them as evidence for a central perturbation of the potential via the ring galaxy mechanism (e.g. Lynds and Toomre 1976). Although the H I velocity field does not support such a simple picture, the existence of the rings is consistent with a disk system in which the gravitational potential is time varying. Such transients might be introduced by an ongoing accretion/merger of a disk system with a lower-mass (say 1/3 mass) companion in which the disk has been disturbed but not quite disrupted (as suggested for other ripple galaxies by Schweizer and Seitzer 1988).

The most likely explanation for Arp 10 is that of a disk system which has been strongly disturbed by a recent large accretion event which has created the ripple-like structures in the outer optical galaxy. Since no large companion is known associated with Arp 10, we suggest that the companion has been almost completely disrupted by the accretion/merger. In the SS88 "ripple galaxies", mass transfer rather than two-body merging is believed to be the dominant mechanism for setting up the ripples. On the other hand, it can only be a matter of time before the companion is so strongly disrupted that its identity is lost. In their sample of ripple galaxies, SS88 show numerous examples of companionless ripples. We suggest that Arp 10 will soon become such a case.

Our earlier optical observations (Figure 1 of CAM) showed that there is a bright extra-nuclear knot between the first and second star forming rings in Arp 10, which we suggested might be in some way related to the nature of the rings. In Figure 8 we show a J–band image of Arp 10 which clearly shows the knot which may be slightly extended in the direction of the nucleus of Arp 10. No obvious H I component was found associated with the knot (except perhaps the peculiar velocity component at 9060 $km\,s^{-1}$ mentioned in Section 4.2) and our optical spectra did not sample the region. Hence, the velocity of the knot is unknown. However, we consider it possible that the knot is a remnant nucleus of a second galaxy which has been disrupted by the collision/merger with the more massive "target". If the collision was close to head-on (in order to set up the rings seen in the inner regions of Arp 10) and the collision was slow, the disruption and merger of the nucleus would be quite rapid (perhaps of the order of two crossing times, or approximately a few × $10^8$ years). The nucleus would fall quickly into the center by the mechanism of tidal friction (Binney and Tremaine 1987) leaving the ripples behind as evidence of a disruptive collision. The fact that the ripples are somewhat asymmetric (the optical ripples are seen mainly in the south) might suggest that only one or two passages of the infalling galaxy have occurred since the initial collision. This would be consistent with the perturbations to the H I velocity field which would be expected to be damped quite quickly as a result of dissipation in the disk.

It is interesting that Arp 10 exhibits both star forming rings similar to those seen in collisional ring galaxies, and faint outer ripples like those seen in the accretion-dominated merger systems. Our observations suggest that when the accreting galaxy is massive enough it can not only produce ripple-like debris, but also drive waves through the disk of the target galaxy. As such, Arp 10 seems to represent an interesting transition case between accretion-driven shell and centrally perturbed disk systems. Very little modeling of this sort of merger has yet been performed. Taniguchi and Noguchi (1991) showed that a collision between two galaxies in which one travels through the center, but co-planar to the disk target disk, can produce shell-like structure and rings not too dissimilar from the structure of Arp 10 (They called these galaxies "Wing" galaxies because of the shape of the debris of the galaxy that was disrupted in the collision). Although Arp 10 is not well represented by this model, the observations do remind us of the huge parameter space that remains unexplored by numerical modeling, especially those involving the disruption of lower-mass galaxies. The observations also underline the importance of obtaining 21cm H I observations, which in this case, led to a different explanation for the origin of the rings from those based on optical observations alone.

Finally, it is worth returning to the horseshoe like structures seen in the H I channel maps of Figure 4 at the most extreme velocities seen in Arp 10. As we discussed in Section 4, these structures are associated with the outer H I disk only. They are not, at present, distinctly separate structures from the rest of the H I disk of Arp 10, but are identifiable only as loops in the velocity-position phase-space of the channel maps. However, it is interesting to note that if the entire **inner** H I disk of Arp 10 was removed (perhaps as a result of the ultimate collapse of the gas towards the center of the system due to cloud–cloud collisions), these loops would then become isolated from the rest of the system. It is perhaps significant that isolated "horseshoe-like" H I emission features, similar to those found in Arp 10, are seen in the outer parts of NGC 2865 and Centaurus A — both shell elliptical systems (Schiminovich et al. 1994, 1995). These structures may therefore be remnants of huge neutral hydrogen disks which preceded the shell-making collisions. If this is true, then those H I loops may contain important information about the progenitors of shell and ripple galaxies[3].

---

[3] We note that if the central regions of a "normal" H I disk were to be removed, the remaining outer filaments of H I would not have the same interesting "horseshoe" character of those found in Arp 10.

## 7. Conclusions

Our optical and H I study of the peculiar galaxy Arp 10 has led to the following conclusions:

- The optical imaging shows evidence for faint filaments and loops suggesting of a merging system similar to the "ripple" galaxies discovered by SS88.

- Optical spectra of the nearby elliptical galaxy, originally suspected of being the colliding companion show it to be a background galaxy.

- The H I emission from Arp 10 shows a disk which extends to 2.7 × the radius of the bright ring. The H I extends outside of the faintest optical features seen in deep CCD images. The most notable feature of the H I emission is that it seems to be composed of two different, but related structures. In the outer regions, the H I "disk" has no clear optical counterpart. Its kinematics suggests distinctly different motions than those found in the inner disk, although the entire H I disk seems to suggest a single coherent structure.

- The inner H I disk shows some correspondence with the optical ring. At least 50% of the ring shows regular rotation, half of the H I ring is either missing or has been severely warped away from the optical ring.

- A simple kinematic model of a rotating disk fails to reproduce the observed velocity field.

- Deviations of the order of 30–50 $\mathrm{km\,s^{-1}}$ (25–30% of the observed velocity field) are found which suggest that the disk is out of equilibrium. Attempts to fit the velocity field with a set of nested rotating and expanding elliptical rings also failed to reproduce the observations.

- The picture most consistent with the observations is that Arp 10 is a disk system which has been strongly disturbed by a recent large accretion of a gas poor galaxy. Such a scenario would explain both the disturbed velocity field of the H I disk and the faint optical "ripples" seen at the outer parts of the system. A possible remnant nucleus of the accreted galaxy is seen near the center of the system.

- The observations of the disturbed, but not completely disrupted H I disk underlines the robustness of large disks to disruption by massive accretion events. Arp 10's H I disk extends out further than the optical ripples and yet has been able to maintain a moderate degree of overall coherence from its outer loopy regions to the inner disk. The galaxy represents an interesting class of merging system which lies intermediate between the classical ring galaxies and the classical shell or ripple systems.

- We note that if the inner part of the H I disk was removed from Arp 10, the properties of the outer loops seen in H I channel maps would strongly resemble features seen in H I maps of shell elliptical systems (Schiminovich et al. 1994, 1995 ). This suggests that some shell-elliptical systems may still contain the remnants of large H I disks similar to the one discussed here. The strong dissipation of angular momentum in the inner disk during a shell-making collision may have caused the gas in the inner regions to fall towards the center, leaving only the outer disk to orbit the remnant. If this is plausible, then Arp 10 is an interesting laboratory for studying the early stages of this transient process.

It would be desirable to obtain high resolution spectroscopy of the possible second nucleus of Arp 10 in order to determine its radial velocity relative to the disk. We note that high resolution VLA observations of the radio continuum emission in Arp 10 have recently been made (Ghigo and Appleton, in preparation) and these contribute further to our understanding of this fascinating galaxy.

We thank C. Struck and R.J. Lavery (Iowa State University), K. Taylor and D. Malin (AAO) for stimulating discussion. The authors enjoyed interesting discussions with J. van Gorkom and D. Schiminovich (Columbia University) in connection with the similarities between Arp 10 and a number of published and unpublished observations of shell elliptical galaxies. We are grateful to E. Brinks (NRAO, Socorro) for useful suggestions during the VLA data reduction process and T. Marston (Drake University) for assistance during the acquisition of the optical spectra. We also thank an anonymous referee for helpful comments about the manuscript. This work is funded under NSF grant AST-9319596.

Table 1
Properties of Arp 10

| | |
|---|---|
| $\alpha$(1950) | $2^{\rm h}15^{\rm m}48.9^{\rm s}$ |
| $\delta$(1950) | $+5°25'26.0''$ |
| Distance | 121 Mpc |
| Inner ring diameter | $5.6'' = 3.2$ kpc |
| Bright ring diameter | $43.0'' = 25.2$ kpc |
| Outer isophotes | $80'' = 46.9$ kpc |
| H I dimensions | $120'' \times 90'' = 70.4 \times 52.7$ kpc |
| $V_{opt}$ | $9160 \pm 30$ km s$^{-1}$ |
| $V_{HI}$ | $9108 \pm 10$ km s$^{-1}$ |
| $\Delta V_{1/2}$ | $249 \pm 15$ km s$^{-1}$ |
| $\Delta V_{1/5}$ | $272 \pm 15$ km s$^{-1}$ |
| $F_H$ | $6.48 \times 10^5$ Jy km s$^{-1}$ |
| $M_H$ | $9.5 \times 10^9$ M$_\odot$ |
| $M_T{}^{\rm a} = \frac{(\Delta V_{1/2})^2 R_{HI}}{G}$ | $4.9 \times 10^{11}$ M$_\odot$ |
| $M_H/M_T$ | 0.019 |

<sup>a</sup>$M_T$ is an estimate of the the total dynamical mass of the galaxy.

Fig. 1.— Greyscale image of Arp 10 through the B–band filter. Note the faint "ripples" at the south and the well defined loop at the northeastern side of the galaxy.

Fig. 2.— The global H I profile of Arp 10. The arrow indicates the systemic heliocentric velocity (9108 km s$^{-1}$) of the galaxy derived from our observations.

Fig. 3.— A contour map of integrated H I distribution. The contour increment is 37.22 Jy beam$^{-1}$ m s$^{-1}$ and the level of the first contour is also 37.22 Jy beam$^{-1}$ m s$^{-1}$. The three open crosses indicate the position of foreground stars and the two solid crosses mark the location of the background elliptical galaxy and the nucleus of Arp 10.

Fig. 4.— Contour plots of the 27 channel maps of Arp 10. The velocity of each channel is displayed in the upper right corner. The contour increment is $4.24 \times 10^{-4}$ Jy beam$^{-1}$ (one $\sigma$ level) and the lowest contour displayed is at $3\sigma$.

Fig. 5.— The velocity structure of the outer (a) and inner (b) regions of Arp 10 superimposed on a deep B–band image. The dark lines indicate the approximate extent of the H I emission ridge-lines or centroids on each channel map.

Fig. 6.— The mean velocity field of Arp 10. The isovelocity contours are at increment of 10 km s$^{-1}$ starting at 9000 km s$^{-1}$ at the north and reaching a peak value of 9220 km s$^{-1}$ at the south. The three open crosses indicate the position of foreground stars and the two solid crosses mark the location of the background elliptical galaxy and the nucleus of Arp 10.

Fig. 7.— The position-velocity diagrams through the center of the galaxy along the major (a) and minor (b) axis. The solid lines show the H I velocities and the black dots show velocities measured in the H II regions using our KPNO spectra.

Fig. 8.— An integrated spectrum of the "companion". We mark the position of Ca II H and K lines, and the G–band.

Fig. 9.— A J–band (1.25 $\mu$m) near-IR image of the central regions of Arp 10. The arrow indicates the extra-nuclear knot which we suspect is the remnant nucleus of a merging second galaxy (Data from 2.1m IRIM photovoltaic array from unpublished work of P. N. Appleton and A. P. Marston)
.